\documentclass[a4paper]{jpconf}

\usepackage{amsmath}
\usepackage{amssymb}
\usepackage{graphicx}
\usepackage{dcolumn}
\usepackage{bm}
\usepackage{booktabs}

\usepackage{subfigure}
\usepackage[utf8]{inputenc}

\graphicspath{{images/}{plots/}}

\newcommand{\avg}[1]{\ensuremath{\left< #1 \right>}}

\newcommand{\brac}[1]{\ensuremath{\left(#1\right)}}

\renewcommand{\vec}[1]{\bm{#1}}

\DeclareMathOperator{\ee}{e}

\begin{document}

    \title{Large Deviations of Convex Hulls of the ``True'' Self-Avoiding Random Walk}
    \author{Hendrik Schawe, Alexander K Hartmann}
    \address{Institut f\"ur Physik, Universit\"at Oldenburg, 26111 Oldenburg, Germany}
    \ead{hendrik.schawe@uni-oldenburg.de}
    \date{\today}

    \begin{abstract}
        We study the distribution of the area and perimeter of the convex hull
        of the ``true'' self-avoiding random walk in a plane. Using a
        Markov chain Monte Carlo sampling method, we obtain the distributions also in
        their far tails, down to probabilities like $10^{-800}$.
        This enables us to test previous conjectures regarding the scaling of
        the distribution and the large-deviation rate
        function $\Phi$. In previous studies e.g., for
        standard random walks, the whole distribution was
        governed by the Flory exponent $\nu$. We confirm this in the present
         study by considering
        expected logarithmic corrections. On the other hand, the behavior of the rate function
        deviates from the expected form. For this exception
        we give a qualitative reasoning.
    \end{abstract}

    \section{Introduction}
        The random walk is a very simple model for diffusive processes with
        Brownian motion~\cite{Einstein1906theorie} as the prime example. Though
        its applications range from financial models~\cite{Fama1965Random}
        over online search engines~\cite{page1999PageRank} to the very sampling
        algorithm used in this study~\cite{newman1999monte}.
        Its simplest variation lives on a lattice and takes steps on random
        adjacent sites at each timestep, which is exceptionally well
        researched~\cite{hughes1996random}.
        With the further constraint that no site may be visited twice, such
        that the walk is \emph{self-avoiding}, it becomes a simple model
        for polymers~\cite{Madras2013}. Interestingly, depending on the exact
        protocol how the self-avoidance is achieved, they can also be used to
        study the perimeter of, e.g., critical percolation clusters~\cite{weinrib1985kinetic}
        or spanning trees~\cite{manna1992spanning,Majumdar1992Exact}

        The distance of a random walk from its starting point is the most prominent
        and simple measurable quantity. Nevertheless, here we go beyond this by considering the
        convex hull of all $T$ sites visited by the random walk, i.e.,
        the smallest convex
        polygon containing all these sites. It can be
        seen as a measure of the general shape of the random walk, without exposing all details
        of the walk. Thus, the area $A$ or perimeter $L$ of the convex hull can then be
        used to characterize the
        random walk in a very simple way. This method is also used, for example,
        to describe the home ranges of animals~\cite{mohr1947,worton1987,boyle2009},
        spread of animal epidemics~\cite{Dumonteil2013spatial} or classification
        of different phases using the trajectory of intermittent stochastic
        processes~\cite{grebenkov2017unraveling}.
        For standard random walks its mean perimeter~\cite{Letac1980Expected} and
        mean area~\cite{Letac1993explicit} in the large $T$ limit are known exactly
        since a long time. More recently different approaches generalized these
        results to multiple random walks~\cite{Majumdar2009Convex,Majumdar2010Random}
        and arbitrary dimensions~\cite{Eldan2014Volumetric}.
        Even more recently the mean perimeter and area for finite (but large)
        walk lengths $T$ were computed explicitly~\cite{grebenkov2017mean} if
        the random walk is discrete-time with jumps from an arbitrary
        distribution. If the distribution of the jump length is Gaussian, even an
        exact combinatorial formula for the mean volume in arbitrary dimensions is
        known~\cite{kabluchko2016intrinsic}. For higher moments however, there
        is only one analytic result for the special case of Brownian
        bridges~\cite{Goldman1996}, i.e., closed walks with Gaussian jumps.
        When asking for more, i.e., for the full distributions,
        no exact analytical results are available. This motivated the numerical
        study of the full distributions---or at least large parts of the
        support---using large-deviation sampling techniques
        to sample even far into the tails of standard random walks~\cite{Claussen2015Convex}
        and multiple random walks~\cite{Dewenter2016Convex}, also in higher
        dimensions~\cite{schawe2017highdim}. These numerical studies are rather
        expensive, since they usually require Markov chain Monte Carlo simulations, allowing one
        to measure the distribution in regions where the probabilities are as small as $10^{-100}$.

        Since self-avoiding walks are considerably more difficult to treat
        analytically than standard random walks, there are no analytical results
        about the properties of their convex hulls yet. Therefore, the authors of this
        contribution very
        recently published a numerical study of the full distribution of
        perimeter and area of three different types of self-avoiding random
        walks~\cite{schawe2018avoiding}, notably the classical \emph{self-avoiding walk} (SAW)
        and the \emph{smart kinetic self-avoiding walk} (SKSAW). While the SAW
        is combinatorial in nature and describes the set of all self-avoiding
        configurations with equal probability, the SKSAW is a growth process,
        which assigns more weight to some configurations.
        In~\cite{schawe2018avoiding} we also give an estimate for the functional form of the rate
        function $\Phi$ describing the far right tail behavior of the distribution,
        i.e., $P(S) \approx \ee^{-T\Phi(S)}$. It was found to depend only on the dimension $d$
        and the scaling exponent $\nu$. For two-dimensional random walks these
        scaling exponents are often known exactly through Schramm-Loewner
        evolution~\cite{cardy2005sle,lawler2002scaling,Lawler2011,Kennedy2015}.

        In this study we test this prediction
        for the \emph{``true'' self-avoiding walk} (TSAW), which has a free parameter
        $\beta$ governing how strictly self-avoiding the walk is.
        Introduced in~\cite{amit1983asymptotic} the TSAW
        was a counter model to the SAW, especially
        it should demonstrate that the behavior of the combinatorial SAW is very
        different from more natural growing random walks which avoid themselves.
        Indeed, in two dimensions, where the end-to-end distance $r$ of a $T$
        step SAW scales as $r \propto T^\nu$ with $\nu=3/4$, the TSAW will
        scale as
        \begin{equation}
            r \propto T^\nu \left(\ln T\right)^\alpha
            \label{eq:end-end}
        \end{equation}
        with $\nu=1/2$~\cite{Pietronero1983critical}
        and a correction $\alpha$, which is not known rigorously, but estimated
        as $\alpha = 1/4$~\cite{Grassberger2017selftrapping}.
        Here we show, for large-enough values of $\beta$,
        that in contrast to previous work \cite{schawe2018avoiding} the rate
        function $\Phi$ is not simply determined by the value of $\nu$, since the growth
        process of the TSAW in the large-area region of the tail is
        indistinguishable from the SKSAW growth process,
        although they have different values of the scaling exponent $\nu$
        determined by the behavior of the high-probability part
        of their distributions.

    \section{Models and Methods\label{sec:mm}}
        This section will introduce the TSAW model and the sampling method in
        enough detail to reproduce the results of this study. For more
        fundamental methods, like the construction of the convex hull, we will
        sketch the main ideas.

    \subsection{Large Deviation Sampling Scheme\label{sec:metropolis}}
\begin{figure}[bhtp]
            \centering
            \includegraphics[scale=1]{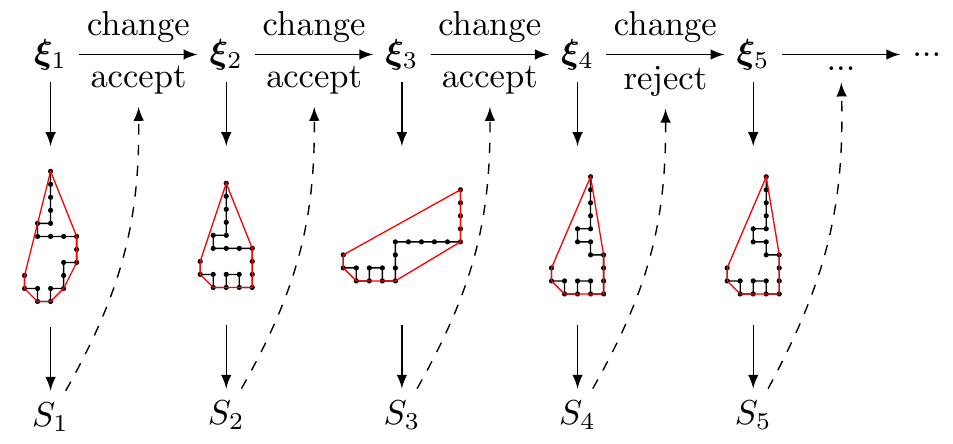}
            \caption{\label{fig:blackBox}
                Sketch of the Markov chain of random number vectors $\vec\xi_i$.
                The change move is performed on the $\vec\xi_i$ and a new TSAW is
                generated from scratch, its energy difference to the previous
                configuration is used to accept or reject the change.
            }
        \end{figure}

        To obtain good statistic in the far tail, it is not sufficient to
        perform naive simple sampling, since configurations of probability $P$
        would need about $1/P$ samples to occur at all. It is therefore not
        feasible to explore the distributions down to the tails of $P < 10^{-100}$
        with simple sampling. Instead we use an importance sampling scheme
        to generate more samples in the low probability tails. Thus, we generate Markov
        chains consisting of sequences of TSAWs and use the
        well known Metropolis algorithm~\cite{metropolis1953equation} with a Boltzmann
        sampling weight. For this purpose,
        we identify the quantity $S$ we are interested in---here the area $A$
        but it could be any measurable quantity in principle---with the energy
        occurring in the Boltzmann factor and
        introduce an artificial ``temperature'' $\Theta$. Since the TSAW is
        a growth process, it is not trivial to come up with a local change move within
        the Markov chain,
        i.e., it is difficult to change a configuration by a small amount
        while preserving the correct statistics. Therefore our Markov chain is
        not directly a chain of configurations of TSAW but rather a chain of random number
        vectors $\vec\xi_i$. Each vector $\vec\xi_i$ determines
        a configuration of a TSAW by performing the growth process and using for
        each of the $T$ decisions a random number from $\vec\xi_i$. This approach
        is sketched in Fig.~\ref{fig:blackBox} and extremely general since it
        can be applied to any model~\cite{Hartmann2014high}. A change move is
        a simple change of one entry of $\vec\xi_i$.

        Following the Metropolis algorithm, we propose a new $\vec\xi^\prime$
        by replacing a random entry with a new random number $\xi \in U[0,1)$,
        generate a new TSAW configuration from scratch using the random numbers
        $\vec\xi^\prime$ and calculating its energy $S^\prime$, i.e., its area.
        The proposed configuration is then accepted, i.e., $\vec\xi_{i+1} = \vec\xi^\prime$,
        or rejected, i.e., $\vec\xi_{i+1} = \vec\xi_i$, depending on the temperature
        and energy difference with respect to the previous configuration with
        probability $p_\mathrm{acc} = \ee^{-\Delta S / \Theta}$, where
        $\Delta S = S^\prime - S_i$ is the energy difference caused by the change.
        Replacing a random entry by a new entry is clearly ergodic, since any
        possible $\vec\xi_i$ can be generated this way. Since we use the classical
        Metropolis acceptance probability, detailed balance is also given.
        This Markov process will therefore yield configurations
        $\vec\xi$ according to their equilibrium distribution
        $Q_\Theta(\vec\xi) = \frac{1}{Z(\Theta)} Q(\vec\xi) \ee^{-S(\vec\xi)/\Theta},$
        where $Q(\vec\xi)$ is the natural, unbiased distribution of configurations
        and $Z(\Theta)$ the corresponding partition function.
        For small temperatures this
        will lead to small energies, i.e., smaller than typical perimeters or
        areas. For large temperatures typical configurations will be generated
        and for negative temperatures larger than usual energies dominate.
        Since this Metropolis algorithm will generate instances following a
        Boltzmann distribution we can easily undo this bias, i.e., we can derive
        the actual distribution $P(S)$ from the biased, temperature dependent
        distributions $P_\Theta(S)$ as
        \begin{align}
            P_\Theta(S) &= \sum_{\{\vec\xi | S(\vec\xi) = S\}} Q_\Theta(\vec\xi)\\
                        &= \sum_{\{\vec\xi | S(\vec\xi) = S\}} \frac{\exp (-S/\Theta)}{Z(\Theta)} Q(\vec\xi)\\
            \label{eq:correction}
                        &= \frac{\exp (-S/\Theta)}{Z(\Theta)} P(S).
        \end{align}

        The unknown $Z(\Theta)$ can be numerically determined by enforcing the
        continuity of the distribution. Therefore we need to simulate the system
        at many different temperatures $\Theta$, such that all histograms $P_\Theta(S)$
        overlap with adjacent temperatures. $Z(\Theta)$ can now be calculated in
        overlapping regions, which should coincide for continuity, i.e.,
        \begin{align}
            \ee^{S/\Theta_i} Z(\Theta_i) P_{\Theta_i}(S) &= \ee^{S/\Theta_{i+1}} Z(\Theta_{i+1}) P_{\Theta_{i+1}}(S)\\
            \Rightarrow \quad \frac{Z(\Theta_i)}{Z(\Theta_{i+1})} &= \exp\left( {S/\Theta_{i+1} - S/\Theta_i} \right) \frac{P_{\Theta_{i+1}}(S)}{P_{\Theta_i}(S)}.
        \end{align}
        This relation fixes all ratios of consecutive $Z(\Theta)$. The absolute
        value can be fixed by the normalization of the whole distribution.

        This method is applicable to a wide range of models, and already
        successfully applied to obtain, e.g., the distributions over a large
        range for the score of sequence alignments~\cite{Hartmann2002Sampling,Wolfsheimer2007local,Fieth2016score},
        work distributions for non-equilibrium systems~\cite{Hartmann2014high},
        properties of Erd\H{o}s R\'enyi random graphs~\cite{Engel2004,Hartmann2011large,hartmann2018distribution},
        and in particular to obtain statistics of the convex hulls of a wide
        range of types of random walks~\cite{Claussen2015Convex,Dewenter2016Convex,schawe2018avoiding}.

    \subsection{``True'' Self-Avoiding Walk\label{sec:tsaw}}
\begin{figure*}[bhtp]
            \centering
            \subfigure[\label{fig:prob:SAW} SAW]{
                \includegraphics[width=0.47\textwidth]{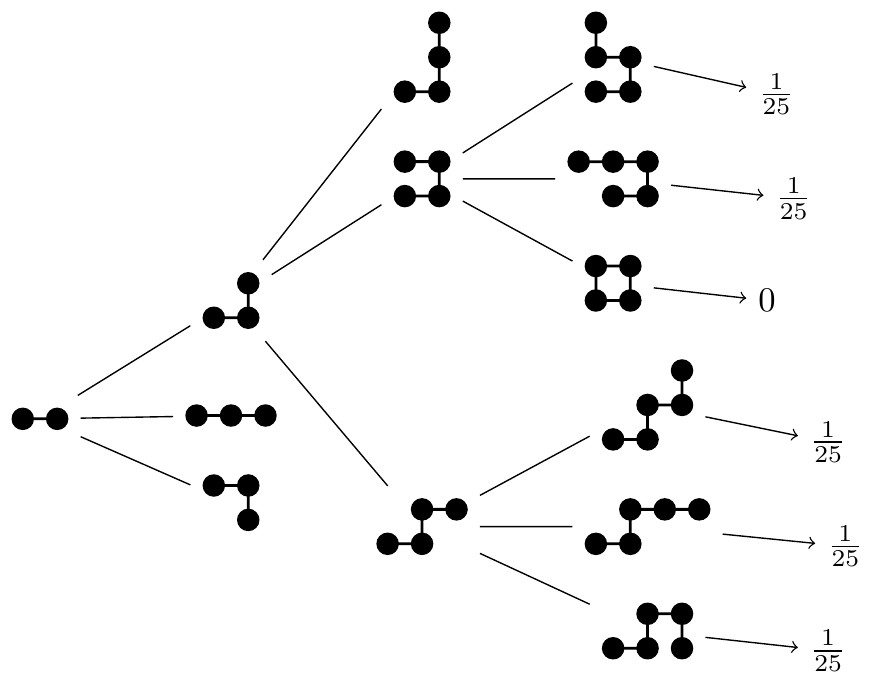}
            }
            \subfigure[\label{fig:prob:TSAW} TSAW]{
                \includegraphics[width=0.47\textwidth]{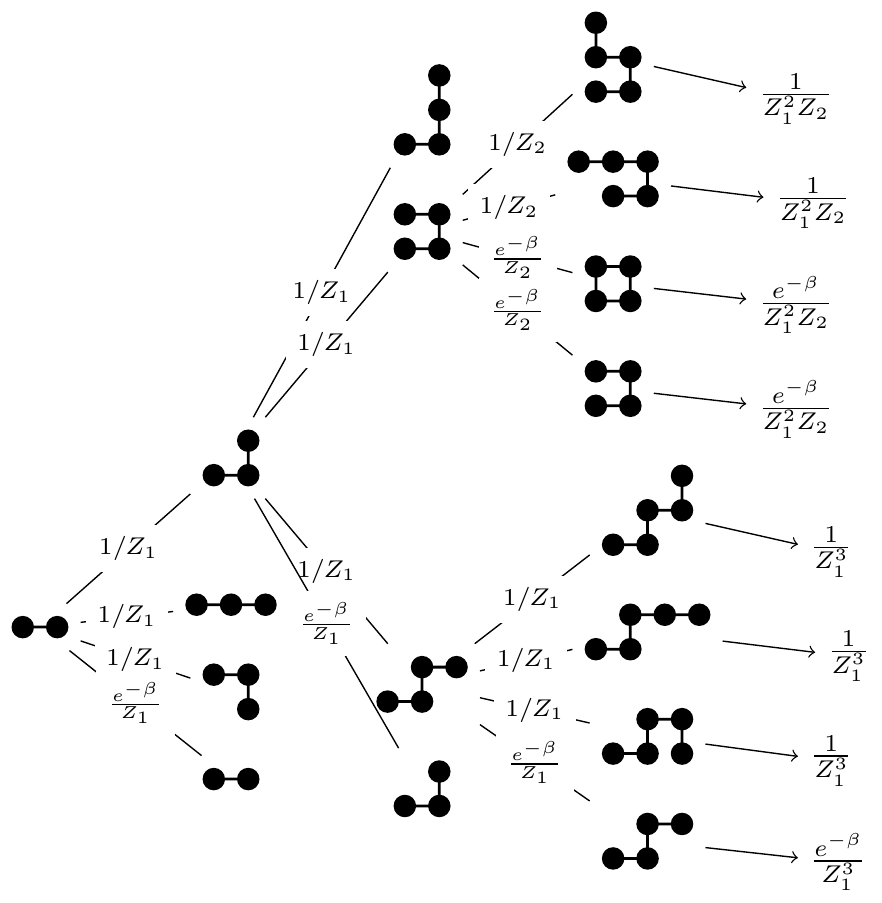}
            }
            \caption{\label{fig:prob}
                Partial decision trees for SAW and TSAW of walks up to length $T=5$.
                On the right side of each tree
                the weight of the configuration is displayed. While the weights
                for the SAW are by definition uniform for every valid configuration,
                the TSAW not only allows self-intersection, but also has
                different weights depending on the history of the walk. Note that $Z_1=3+\exp(-\beta)$
                and $Z_2=2+2\exp(-\beta)$.
            }
        \end{figure*}

        The ``true'' self-avoiding Walk (TSAW) is a random walk model, in which
        the walker tries to avoid itself, but self-avoidance is not strictly imposed.
        \begin{figure*}[bhtp]
            \centering

            \subfigure[\label{fig:tsaw:b0} $\beta = 0$]{
                \includegraphics[scale=0.7]{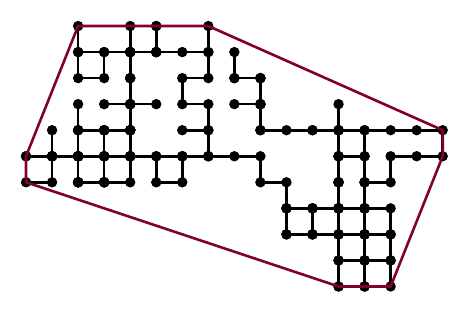}
            }
            \subfigure[\label{fig:tsaw:b1} $\beta = 1$]{
                \includegraphics[scale=0.7]{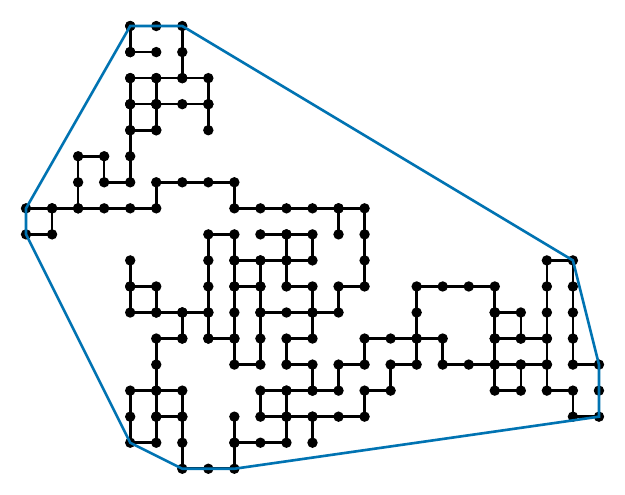}
            }
            \subfigure[\label{fig:tsaw:b100} $\beta = 100$]{
                \includegraphics[scale=0.7]{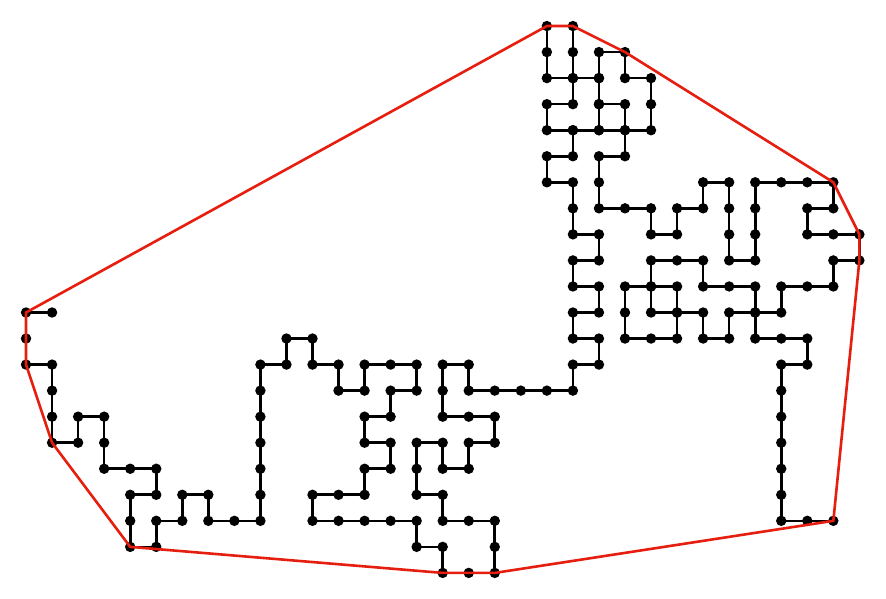}
            }
            \caption{\label{fig:tsaw}
                Examples of typical TSAW realizations at different values of
                the avoidance parameter $\beta$ with their convex hulls.
                Each walk has $T=200$ steps. Larger values of $\beta$ lead
                to larger extended walks characterized by larger areas of
                their convex hulls.
            }
        \end{figure*}
        To construct a TSAW realization one
        grows a standard random walk on a lattice and records the number of
        visits $n_i$ to each site $i$. For each step the probability to
        step on a neighboring site $i$ is weighted with the number of times that
        site was already visited
        \begin{align}
            p_i = \frac{\exp\brac{-\beta n_i}}{\sum_{j\in \mathcal{N}} \exp\brac{-\beta n_j}},
        \end{align}
        where the sum over all current neighbors $\mathcal{N}$ is for normalization.
        The free parameter $\beta$ governs the strength of the avoidance. Large
        values of $\beta$ lead to stronger avoidance, negative values of
        $\beta$ lead to attraction and $\beta=0$ is the special case of the
        standard random walk. For a selection of $\beta$ values typical
        examples are visualized in Fig.~\ref{fig:tsaw}.
        The TSAW is not to be confused with the classical
        self-avoiding walk (SAW), which describes the ensemble of all
        configurations which do not intersect themselves each weighted
        the same. In Fig.~\ref{fig:prob}
        two partial decision trees are displayed which visualize the fundamental
        differences in the weights of the configurations. Even in the
        $\beta\to\infty$ limit ($Z_1=3$, $Z_2=2$) the weights differ.
        In particular its
        upper critical dimension is $d=2$~\cite{amit1983asymptotic}, which
        means that the exponent $\nu$, which characterizes the scaling of the end-to-end distance
        $r \propto T^\nu$, is $\nu = 1/2$ with logarithmic corrections,
        i.e., $r \propto T^{\nu} \brac{\ln T}^\alpha$, where $\alpha = 1/4$~\cite{Grassberger2017selftrapping}
        is conjectured.

    \subsection{Convex Hulls\label{sec:convex}}
        The convex hull of a set of points $\mathcal{P}$ in the plane is the
        smallest convex polygon enclosing every point $p \in \mathcal{P}$ and
        hence also every line between any pair of points. Some examples of
        convex hulls are visualized in Fig.~\ref{fig:tsaw}. The construction
        of a convex hull of a planar point set is a solved problem, in the sense
        that an optimal algorithm exists~\cite{Kirkpatrick1986ultimate,Chan1996Optimal}
        with result-dependent run time $\mathcal{O}(T\log h)$, where $T$ is the number of points
        $|\mathcal{P}|$ and $h$ is the number of vertices of the resulting
        convex hull. In practice, however, suboptimal but simpler and for point
        sets as small as in this study ($T \approx 10^6$) faster algorithms are
        used. Especially for planar point sets one can exploit the fact that
        a polygon can be defined by the order of its vertices, instead by a
        list of its facets. The Graham scan~\cite{Graham1972Efficient} algorithm
        is based on this fact. After shifting the coordinate origin into the
        center of the point set, it sorts the points according to their polar
        coordinate. Then starting at an arbitrary point all points are filtered
        out which are oriented clockwise with respect to the the previous
        and next (not-filtered out) points. Iterating this over a full revolution,
        leaves only the points which constitute the vertices of the convex hull.
        This algorithm is dominated by the time to sort the points, which can
        be done in $\mathcal{O}(T\log T)$. Here, we use Andrew's monotone chain
        algorithm~\cite{Andrew1979Another}, which is a variation of the Graham scan
        sorting the points lexicographically, which is slightly faster, instead of by
        polar angle. Note that this type of algorithm does not generalize to
        $3$ or higher dimensions. For those
        cases a different algorithm, like quickhull~\cite{Barber1996thequickhull}
        has to be used. Before applying the exact algorithm, we reduce the size
        of the point set with Akl's elimination heuristic~\cite{Akl1978Fast},
        which removes all points inside, in our implementation, a octagon of
        extreme points. Of a few tested polygons the octagon showed the best
        performance in the instances we typically encounter in this study.

        To calculate the area $A$ of a convex polygon, where the coordinates are
        sorted counterclockwise, one can sum the areas of the trapezoids
        extending perpendicular to the $x$-axis
        \begin{align}
            A &= \frac{1}{2}\sum_{i=0}^{h-1} (y_i+y_{i+1}) (x_i - x_{i+1}).
        \end{align}
        The perimeter $L$ is the sum of the line segments of consecutive points
        of the hull
        \begin{align}
            L &= \sum_{i=0}^{h-1} \sqrt{(x_i - x_{i+1})^2 + (y_i+y_{i+1})^2},
        \end{align}
        with $x_h \equiv x_0$ and $y_h \equiv y_0$.

    \section{Results}
        We simulated the TSAW at two values of $\beta$. The limit case of a
        TSAW, which only steps on itself, if it has no other choice, was
        simulated at $\beta=100$. Since the probability to step on already
        visited sites is exponential in $\beta$, this corresponds to the
        $\beta\to\infty$ case. Further, we simulated at $\beta=1$, to capture
        also the case, which does sometimes voluntarily step on itself.

\begin{figure*}[bhtp]
            \centering
            \subfigure[\label{fig:means} Asymptotic mean values]{
                \includegraphics[scale=1]{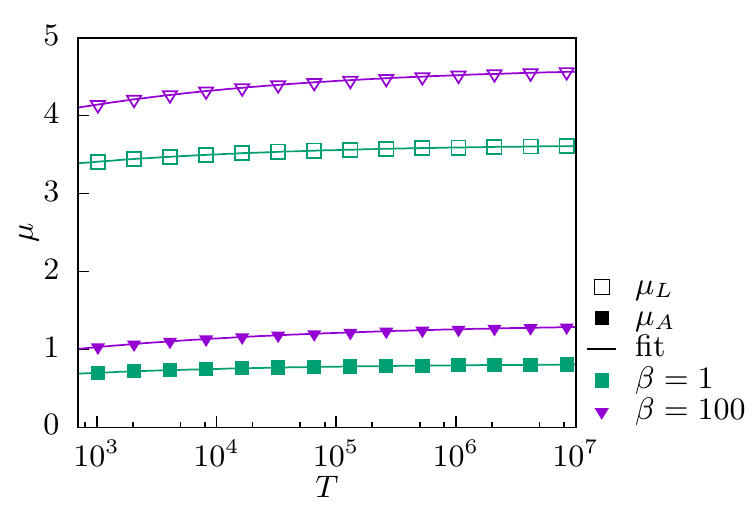}
            }
            \subfigure[\label{fig:compare} Comparison of different types of random walks]{
                \includegraphics[scale=1]{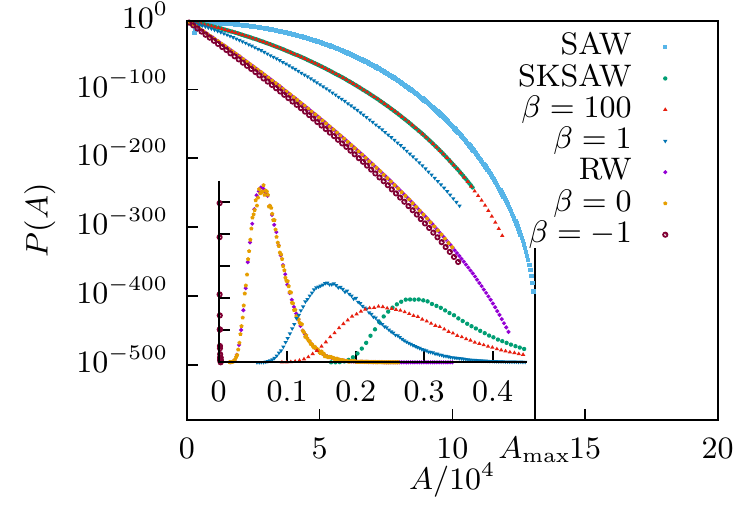}
            }
            \caption{\label{fig:combined}
                \subref{fig:means} Extrapolation of the asymptotic mean values
                of the perimeter and area of the convex hull.
                \subref{fig:compare} Distribution of all scrutinized walk types with $T=1024$ steps.
                The inset shows the peak region. Note that the standard random
                walk (RW) and the $\beta=0$ TSAW coincide. Also the far right
                tail of the $\beta = 100$ TSAW and SKSAW coincide, but not the
                main region. The vertical line shows the maximum area constructable
                with $1024$ steps, which has an area $A_{\max}$ about
                $512^2/2\approx 1.3\times 10^5$. The distributions are thus not
                sampled over their whole support, but a large region.
            }
        \end{figure*}

        First, we will look at the behavior of the mean of the perimeter and
        area. Here, we used simple sampling for walk lengths in the range
        $T\in\{2^k | 10 \le k \le 23\}$. Each value is averaged over $10^6$ TSAWs.
        Naturally, the mean of geometric volumes scale with their intrinsic dimension
        $d_i$ and a typical length scale $r$, e.g., the end-to-end distance,
        as $r^{d_i}$. Using the scaling of $r$ from Eq.~\eqref{eq:end-end}, we expect the mean values of
        the perimeter $\avg{L}$ ($d_i=1$ in $d=2$) and the area $\avg{A}$ ($d_i=2$)
        to scale as
        \begin{align}
            S \propto T^{d_i\nu} \ln(T)^{d_i\alpha}
            \label{eq:scale_mean}
        \end{align}
        for large values of $T$. We can even
        calculate the asymptotic prefactors $\mu^\infty$ by extrapolating the
        scaled values for finite sizes $\mu_L=\avg{L} T^{-1/2} \ln(T)^{-\alpha}$ and
        $\mu_A=\avg{A} T^{-1} \ln(T)^{-2\alpha}$ to their asymptotic values
        $\mu_L^\infty$ and $\mu_A^\infty$. For the extrapolation, which is shown
        in Fig.~\ref{fig:means}, we use a simple power law with offset $\mu = \mu^\infty - a T^{-b}$,
        which were already used for this purpose in \cite{Claussen2015Convex,Dewenter2016Convex}.
        The asymptotic values are listed in table~\ref{tab:mean}. As expected
        the values for the TSAW are larger for larger $\beta$. To our knowledge,
        there are no analytical calculations for these asymptotic values to which we could
        compare to. The given
        error estimates are only statistical and do not include the systematic
        error introduced by the ad-hoc fit function. Nevertheless the convergence
        of the values is very well visible, confirming $\nu=1/2$ and $\alpha=1/4$
        to be very good estimates.

        Direct fits of the form Eq.~\eqref{eq:scale_mean} yield values in good
        agreement with the expected exponents for the end-to-end distance $r$
        at $\beta = 1$, but most other data sets lead to fits overestimating
        $\alpha$ and slightly underestimating $\nu$. A possible, at least partial,
        explanation for this is be that the relation $L(r)$ is not perfectly
        linear for the sizes we obtained data for.

        \begin{table}[htb]
            \centering
            \begin{tabular}{l c c}
                \toprule
                                         & $\mu_L^\infty$ & $\mu_A^\infty$ \\
                \midrule
                TSAW $\beta = 0$ (exact) & $3.5449...$ & $0.7854...$ \\
                TSAW $\beta = 1$         & $3.636(2)$  & $0.820(1)$ \\
                TSAW $\beta = 100$       & $4.641(3)$  & $1.339(3)$ \\
                \bottomrule
            \end{tabular}
            \caption{
                Asymptotic mean values of the area and perimeter scaled by Eq.~\eqref{eq:scale_mean}.
                The values are obtained by the fit shown in Fig.~\ref{fig:means}.
                The given error estimates are only statistical and do not
                include the systematic error introduced by the ad-hoc fit function.
                The exact values are from \cite{Eldan2014Volumetric} and converted
                to a square lattice as described in \cite{schawe2018avoiding}.
            }
            \label{tab:mean}
        \end{table}

        We now focus on the main result, on the distribution $P(A)$ of the
        convex-hull area. These results were obtained using the large-deviation Markov-chain
        simulations. We had to perform simulations for different ``temperatures''
        ranges for each system size and parameter $\beta$. For example $T=128$ at $\beta=1$
        needed seven temperatures for the right tail $\theta \in [-40,-9]$ and three
        more for the left $\theta \in [7,40]$. For larger system sizes more
        temperatures are usually needed. For the $\beta=1$ case at $T=2048$ we
        used 32 temperatures $\theta \in [-3200,-105]$ to obtain the right tail.
        For the $\beta=100$ cases we could use similar values for the temperatures.
        Equilibration was ensured as described in \cite{schawe2018avoiding}.
        In Figure~\ref{fig:compare} we compare distributions
        of different random
        walk types with the result for the TSAW at different values of $\beta$.
        By using the large-deviation algorithm, we were able to obtain this distribution
        over hundreds of decades in probability, down to values as small as $P(A)\sim 10^{-800}$
        for the largest value of $T$.
        Notice that while SKSAW and TSAW with high values of $\beta$ show the
        same behavior in the far tail, where the walks are so stretched out
        such that trapping does not play a role anymore. In the main region
        however, they are clearly distinct, as is expected due to their
        different scaling exponent $\nu$. Further, the parameter $\beta$ can
        apparently be used to interpolate the tail behavior between the standard
        random walk case and the SKSAW case.

\begin{figure*}[bhtp]
            \centering
            \subfigure[\label{fig:scaling:beta1} $\beta = 1$]{
                \includegraphics[scale=1]{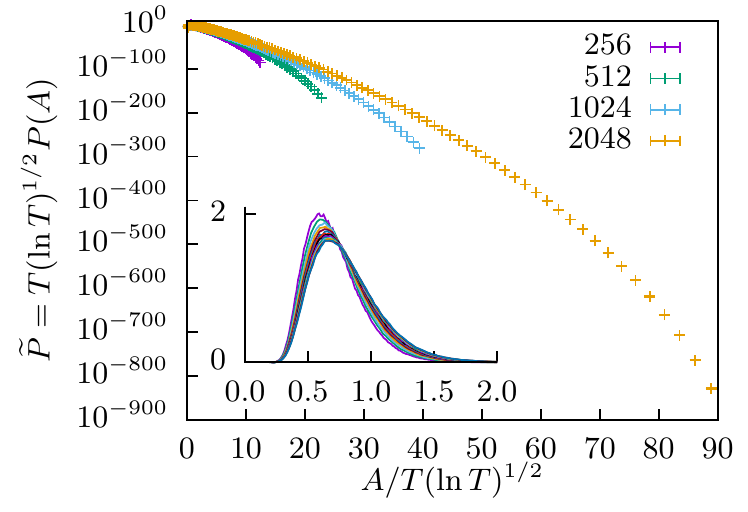}
            }
            \subfigure[\label{fig:scaling:beta100} $\beta = 100$]{
                \includegraphics[scale=1]{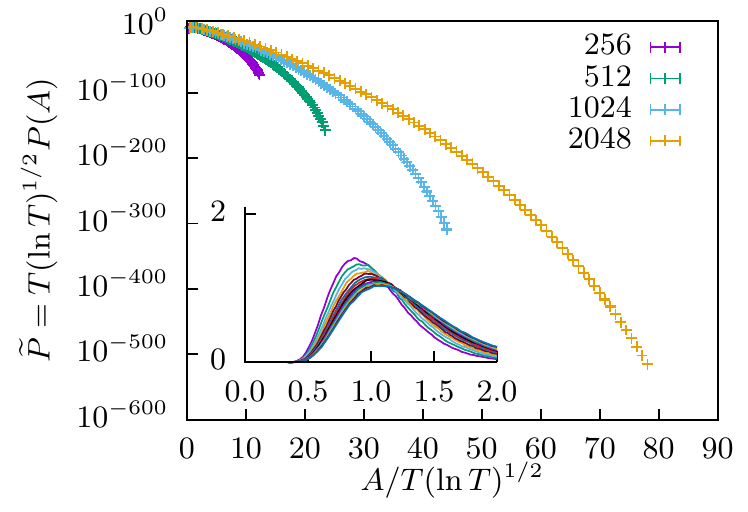}
            }
            \caption{\label{fig:scaling}
                Distributions of the area of the ``true'' self-avoiding random
                walk scaled according to Eq.~\eqref{eq:scaling} plus logarithmic
                corrections for different values of $\beta$ and lengths $T$.
                The insets show the main region for 14 values of $T \in \{2^k | 10 \le k \le 23\}$
                obtained by simple sampling.
            }
        \end{figure*}

        Since we have obtained large parts of the distribution, it would be
        interesting if the whole distribution scales the same as the mean
        values (cf.~Eq.~\eqref{eq:scale_mean}).
        For other types of walks, the distribution of perimeter and area
        could indeed be scaled
        \cite{Claussen2015Convex,Dewenter2016Convex,schawe2017highdim,schawe2018avoiding}
        across their full support only knowing $\nu$, as
        \begin{align}
            P(S) = T^{-d\nu} \widetilde{P}\brac{ST^{-d\nu}}.
            \label{eq:scaling}
        \end{align}
        For the TSAW, this collapse, when considering the logarithmic corrections as visualized
        in Fig.~\ref{fig:scaling}, exhibits an apparent drift towards a limiting shape.
        Nevertheless, severe finite size effects are visible, especially
        in the tails but also in the main region. Despite far larger system sizes $T$
        considered, here the main region collapse is worse than for other kinds
        of self-avoiding walks as shown in \cite{schawe2018avoiding}.
        The stronger finite size effect may be caused by the fact that all walks start on
        an empty lattice. This means for our case that the first steps
        of the walk behave differently from the last steps of the
        walk, when many sites are occupied. Although for the limit of large system
        sizes $T$, the latter should determine the behavior. A possible
        improvement to simulate TSAWs is suggested in~\cite{Grassberger2017selftrapping},
        which is to simulate a much longer walk with $t \gg T$ steps and
        look at the last $T$ steps.

        The rate function $\Phi$ is defined if the distribution obeys the
        \emph{large deviation principle}. This means that  the
        distribution, for large values of $T$, should decay exponentially in the length $T$
        scaled by the rate function as
        \begin{align}
            P_T(s) \approx \ee^{-T\Phi(s)}.
            \label{eq:def_rate}
        \end{align}
        Usually the parameter $s$ is between $0$ and $1$. We achieve this by
        dividing our measured area by the maximum area, i.e., by
        measuring $s = \frac{A}{A_\mathrm{max}}$.
        In two dimensions the walk configuration with maximum area is L-shaped with
        arms of equal length (for odd $T$) and therefore
        $A_\mathrm{max} = \frac{1}{2}\left(\frac{T+1}{2}\right)^2 \approx \frac{T^2}{8}$.

\begin{figure*}[bhtp]
            \centering

            \subfigure[\label{fig:rate:beta1}]{
                \includegraphics[scale=1]{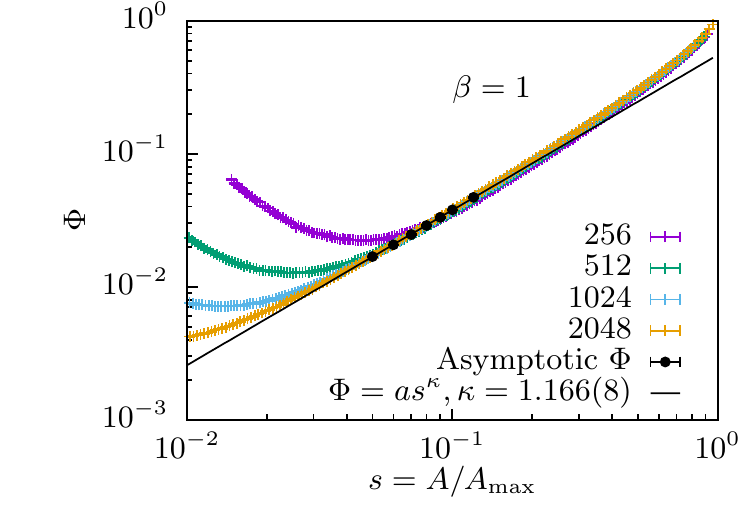}
            }
            \subfigure[\label{fig:rate:beta100}]{
                \includegraphics[scale=1]{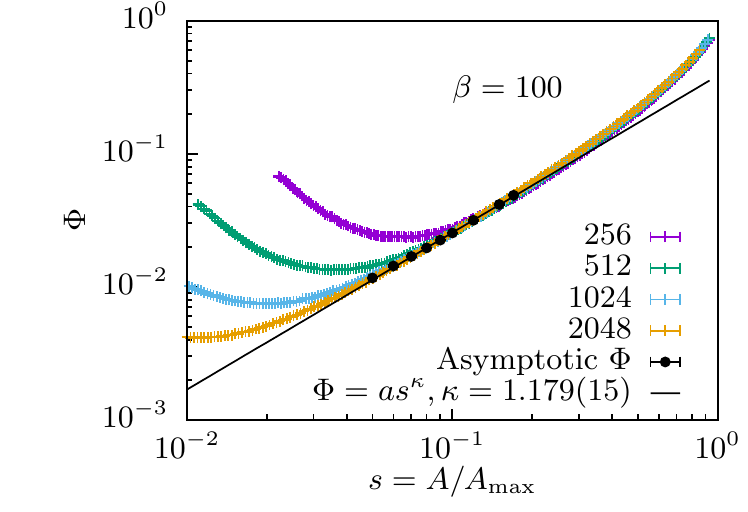}
            }
            \caption{\label{fig:rate}
                Rate function $\Phi$ for the area with fits to the assumed power-law form.
                The fit is performed over a range, where the finite-size influence
                of the lattice should be small, but the large $T$ behavior can
                be extrapolated. Finite-size effects seem more pronounced for
                larger values of $\beta$.
            }
        \end{figure*}

        Similar to~\cite{schawe2018avoiding} we assume the rate function to be
        a power law
        \begin{align}
            \Phi(s) \propto s^{\kappa},
            \label{eq:power-law}
        \end{align}
        which seems to agree reasonably well with our data, since the double
        logarithmic plot Fig.~\ref{fig:rate} shows that the rate function appears
        as a straight line in the intermediate tail. The far tail is dominated
        by finite-size effects caused by the lattice structure, which leads to
        a ``bending up'' of the rate function. For small values of $s$, in the high-probability
        region, the rate
        function does not have any relevance.
        Assuming that the rate function is a power law Eq.~\eqref{eq:power-law}
        and scaling of the form Eq.~\eqref{eq:scaling}
        is possible, with $d_i$ being the intrinsic dimension of the observable, e.g., $d_i=2$
        for the area, we can derive a value for the power law exponent of the
        rate function $\kappa$.
        Using the definition of the rate function Eq.~\eqref{eq:def_rate}
        \begin{align}
            \ee^{-T\Phi(ST^{-d_i})} \approx T^{-d_i\nu} \widetilde{P}\brac{ST^{-d_i\nu}}
        \end{align}
        should hold in the right tail. The $T^{-d_i\nu}$ term can be ignored next to the exponential,
        also the logarithmic correction is subdominant and would not allow to add any insight.
        Apparently the right hand side is a function of $ST^{-d_i\nu}$, such that
        the left hand side also has to be a function of $ST^{-d_i\nu}$. This is the
        case for~\cite{schawe2018avoiding}
        \begin{align}
            \kappa = \frac{1}{d_i(1-\nu)}.
        \end{align}

        To test whether the results for the rate function in case of the TSAW
        obeys this relation, we estimate
        the value of $\kappa$ from our data. Since we have data for various values of the
        walk length $T$, we first
        extrapolate our data point-wise to large $T$. For this purpose we fit a power law with
        offset, where the parameters depend on the value of $s$:
        \begin{align}
            \Phi(s,T) = a(s) T^{b(s)} + \Phi_\infty(s).
            \label{eq:extrapolate}
        \end{align}
        This results in the value of interest $\Phi(s)\equiv \Phi_\infty(s)$, the
        parameters $a(s)$ and $b(s)$ are only auxiliary quantities.
        We perform this extrapolation in a region which is far away from the
        finite-size effects of the far tail. In this range of medium values of $s$
        the extrapolation according
        to Eq.~\eqref{eq:extrapolate} works robustly. Since the bins of the
        logarithmic histograms we use do not have the same borders for every
        system size $T$, we have a-priory not access to the same value of $s$ for
        different values of $T$. Thus,
        we use cubic splines to interpolate such that we obtain results for the same
        value of $s$ for all walk lengths.
        We found cubic splines to be sufficient since the bins are
        rather dense such that systematic errors introduced by the interpolation
        should be small. The final values $\Phi(s)$ obtained from the extrapolated
        values to the assumed form Eq.~\eqref{eq:power-law} are shown in
        Fig.~\ref{fig:rate} as symbols. Next, we fit power laws to this data.
        The values for $\kappa$ obtained are
        within errorbars consistent with $\kappa = 7/6$ which is the expected
        value for the SKSAW ($\nu = 4/7$) and incompatible with the expected
        value $\kappa = 1$ of $\nu = 1/2$ walks.
        This behavior is nevertheless plausible since in the tail (large-area) region, structures,
        which enable self trapping, i.e., loops, are rare since they do lead to
        smaller areas of the convex hull than straight regions. Therefore the
        influence of trappings should diminish in the large area tail, which is
        the main difference in the behavior of SKSAW and TSAW. Without trappings
        the TSAW in the $\beta \to \infty$ limit is functionally identical to
        the SKSAW. Apparently already $\beta=1$ is large enough to produce this
        behavior. Therefore it is natural that the large-area tail behaves the
        same as the SKSAW. On the other hand, to possibly see a range where
        the rate function behaves like a power law with $\kappa=1$ according
        to $\nu=1/2$, one would have to go to much larger system sizes, because
        one would have to obtain data to the right of the peak, but for very
        small values of $s=A/A_{\max} \ll 10^{-2}$, where trappings still do
        play a role. In particular the analysis might be hampered by the
        presence of the logarithmic correction to the mean end-to-end distance.

        This means that the TSAW is more complex in comparison to some other
        types of self-avoiding walks for which it was possible to predict the
        tail behavior from the same exponent $\nu$ which predicts the mean
        behavior. The other types of random walk were under scrutiny in~\cite{schawe2018avoiding},
        namely the self-avoiding walk, the loop-erased random walk and the smart
        kinetic self-avoiding walk (SKSAW). Instead for TSAW in the large
        deviation region a different scaling exponent seems to hold, which is
        very close to the scaling exponent of the SKSAW.

    \section{Conclusions\label{sec:conclusion}}
        We studied the behavior of the distribution of the area of the convex hull of
        the ``true'' self-avoiding walk, especially in the large deviation
        regime of larger than typical areas. With a sophisticated large-deviation
        sampling algorithm, we obtained the distribution over a large part of
        its support down to probabilities smaller than $10^{-800}$ for a typical
        avoidance parameter of $\beta = 1$ and a large avoidance $\beta = 100$.
        The distributions seem to approach a limiting scaling form when rescaled by the
        behavior of the mean, but with much stronger finite-size effects as compared
        to other types of random walks, which were previously studied.

        Using this data we calculated the rate functions. The rate function seem also to
        behave qualitatively similar in comparison to other types of self-avoiding walks
        studied earlier~\cite{schawe2018avoiding} in that they seem to be well
        defined and well approximated by a power law. In contrast to other
        types of random walks, this power law can apparently not be derived
        from the scaling exponent of the mean values $\nu$. Instead it seems
        that a second exponent governs the scaling behavior of the tail for the
        TSAW, which is close to $4/7$, the scaling exponent of the smart kinetic
        self-avoiding walk. This is plausible since the large-area region
        should be dominated by configurations in which no trappings are
        possible, which is the major difference between these types.

        Finally, we also provided estimates for the relevant scale factors of the
        mean of area and perimeter of the convex hulls of TSAWs. They might be
        accessible to analytic calculations in the future.

        For future numerical work it would be interesting to look for further types of
        random walks, which show similar effects of distinct scaling exponents
        for different parts of the distribution but do not show the strong
        logarithmic corrections to scaling, and would therefore be easier to
        analyze. On the other hand, it would be very exciting if one was able to
        obtain data in the range where the rate function exhibits the exponent $\kappa(\nu=1/2)=1$,
        with the need to simulate really large system sizes, but closer to the typical behavior.

    \ack
        This work was supported by the German Science Foundation (DFG) through
        the grant HA 3169/8-1.
        We also thank the GWDG (G\"ottingen) for providing computational resources.

    \section*{References}
        \bibliographystyle{iopart-num}
        \bibliography{lit}

\end{document}